\documentclass[reprint,superscriptaddress,nobibnotes,nofootinbib,amsmath,amssymb,aps,hidelinks]{revtex4-2}
\usepackage{showyourwork}
\usepackage[version=4]{mhchem}
\usepackage{graphicx}
\usepackage{bm}
\usepackage{siunitx}
\usepackage{booktabs}
\DeclareSIUnit{\angstrom}{\textup{\AA}}
\newcommand{\D}{D^*}
\newcommand{\pr}[2]{p(#1\,|\,#2)}
\newcommand{\Ea}{E_{\mathrm{a}}}
\newcommand{\papertitle}{Bayesian Methods for the Investigation of Temperature-Dependence in Conductivity}
\usepackage[noabbrev,nameinlink,capitalize]{cleveref}
\crefname{equation}{Eqn.}{Eqns.}
\crefformat{equation}{Eqn.~#2#1#3}
\crefrangeformat{equation}{Eqns.~#3#1#4 to~#5#2#6}
\crefmultiformat{equation}{Eqns.~#2#1#3}{ and~#2#1#3}{, #2#1#3}{ and~#2#1#3}
\crefname{figure}{Fig.}{Figs.}

\makeatletter
\def\maketitle{
\@author@finish
\title@column\titleblock@produce
\suppressfloats[t]}
\makeatother

\begin{document}

\let\oldaddcontentsline\addcontentsline
\renewcommand{\addcontentsline}[3]{}

\title{\papertitle}

\author{Andrew R. McCluskey}
\email{andrew.mccluskey@bristol.ac.uk}
  \affiliation{Centre for Computational Chemistry, School of Chemistry, University of Bristol, Cantock's Close, Bristol, BS8 1TS, UK.}
  \affiliation{Diamond Light Source, Harwell Campus, Didcot, OX11 0DE, UK.}
\author{Samuel W. Coles}
  \affiliation{Yusuf Hamied Department of Chemistry, University of Cambridge, Lensfield Road, Cambridge, CB2 1EW, UK.}
  \affiliation{Lennard-Jones Centre, University of Cambridge, Trinity Lane, Cambridge, CB2 1TN, UK.}
\author{Benjamin J. Morgan}
\email{b.j.morgan@bath.ac.uk}
  \affiliation{Department of Chemistry, University of Bath, Claverton Down, Bath, BA2 7AY, UK.}
  \affiliation{The Faraday Institution, Quad One, Harwell Science and Innovation Campus, Didcot, OX11 0RA, UK.}

\begin{abstract}
Temperature-dependent transport data, including diffusion coefficients and ionic conductivities, are routinely analysed by fitting empirical models such as the Arrhenius equation. 
These fitted models yield parameters such as the activation energy, and can be used to extrapolate to temperatures outside the measured range.
Researchers frequently face challenges in this analysis: quantifying the uncertainty of fitted parameters, assessing whether the data quality is sufficient to support a particular empirical model, and using these models to predict behaviour at temperatures outside the measured range. 
Bayesian methods offer a coherent framework that addresses all of these challenges. 
This tutorial introduces the use of Bayesian methods for analysing temperature-dependent transport data, covering parameter estimation, model selection, and extrapolation with uncertainty propagation, with illustrative examples from molecular dynamics simulations of superionic materials.
\end{abstract}

\maketitle

\section{Introduction}
\label{sec:intro}

Transport coefficients, such as the self-diffusion coefficient $D^*$ and ionic conductivity $\sigma$, are critical parameters for characterising the performance of technologies such as batteries, fuel cells, and memristors.
These coefficients are typically temperature dependent, with this temperature dependence commonly described by fitting empirical models to the measured data.
Fitted models offer several benefits over raw data alone: they allow interpolation to reveal trends that otherwise would be obscured by noise, extrapolation of transport coefficients to outside the measured regime, and extraction of parameters, such as activation energies, that can be used to compare different materials or to explain microscopic dynamics~\cite{Canepa2022, Wang2021}.

The most widely used empirical model for describing the temperature dependence of transport coefficients is the Arrhenius equation.
For diffusion coefficients, this takes the form
\begin{equation}
    \D = A\exp\left(\frac{-\Ea}{RT}\right),
    \label{eqn:arrhenius_diff}
\end{equation}
where $\Ea$ is the activation energy and $A$ is the pre-exponential factor.
For conductivity, the equivalent expression is\footnote{The Nernst-Einstein relation between $\sigma$ and $\D$ introduces a $T^{-1}$ factor, so the Arrhenius form applies to $\sigma T$ rather than $\sigma$.}
\begin{equation}
    \sigma T = A\exp\left(\frac{-\Ea}{RT}\right),
    \label{eqn:arrhenius}
\end{equation}
Both equations can be linearised by taking logarithms, and it is common to plot the natural logarithm of $D^*$ or $\sigma T$ against inverse temperature: this representation is known as an ``Arrhenius plot'', and data that approximate a straight line are said to exhibit Arrhenius behaviour.

%

%

While many systems show approximate Arrhenius behaviour, this is not universal. 
For some systems, Arrhenius plots show pronounced curvature, corresponding to ``non-Arrhenius'' behaviour that is not well-modelled by the Arrhenius equation. 
In these cases, alternative models---such as the Vogel-Tammann-Fulcher equation~\cite{Vogel1921, Fulcher1925, Tammann1926, deSouza2006, Wang2021}---may provide a better description. 
However, it is not always clear whether observed deviations indicate genuine non-Arrhenius behaviour or simply reflect noise in the data, raising the question of which model is best supported by the data.

Fitting an empirical model requires determining the values of its parameters, e.g., for the Arrhenius model, the activation energy $\Ea$ and pre-exponential factor $A$.
Because the underlying data have scatter or associated uncertainties, we need not only best-fit parameter values but also uncertainty estimates---known as inverse uncertainties---for those fitted parameters.

Once a model has been fitted, a common use is to extrapolate to temperatures outside the measured range; for example, to predict room-temperature conductivity from high-temperature simulation data~\cite{Yeandel2019,Goldmann2025}.
Such extrapolations assume the model remains valid in the new regime. Additionally, for the extrapolated estimates to be meaningful, the uncertainties in the fitted parameters must propagate into the predicted values; without them, we have no way to judge how reliable the extrapolation is. 

These challenges---obtaining fitted parameters with meaningful uncertainties, assessing which model is best supported by the data, and quantifying uncertainty in extrapolations---are non-trivial and often not addressed by standard fitting approaches. Bayesian methods provide a coherent framework for addressing all three: they yield full probability distributions for fitted parameters, enable rational comparison of competing models, and allow uncertainty propagation into extrapolated predictions.

This tutorial discusses the application of Bayesian methods to the analysis of temperature-dependent transport data. 
Our examples focus on conductivity estimates from molecular dynamics simulations of superionic materials, but the methodology is generally applicable to any temperature-dependent data with associated uncertainties. 
The methods outlined here are available in the open-source Python package \textsc{kinisi}~\cite{McCluskey2024}.

\section{Challenges in Temperature-Dependent Modelling}
\label{sec:problems}

Fitting empirical models to temperature-dependent transport data raises three significant challenges: selecting an appropriate model, determining parameter values with meaningful uncertainties, and extrapolating reliably to unmeasured temperatures. 
We examine each of these challenges below, before showing how Bayesian methods address them in Sections \ref{sec:parameter_estimation}--\ref{sec:extrapolation}.

\subsection{Model Selection}

Given a set of temperature-dependent data, how do we determine which empirical model best describes the underlying behaviour? 
If we fit multiple candidate models and compare residuals, a more complex model with more free parameters will generally fit better than a simpler one, even when the additional complexity is not justified by the data. 
Using an overly complex model risks fitting noise rather than signal, producing parameter estimates that are precise but inaccurate, and extrapolations that are unreliable.

A model with more free parameters can fit a wider range of possible behaviours. 
If such a model fits the data, this is less informative than if a simpler, more constrained model fits equally well. 
The simpler model has made a more specific prediction that happened to be confirmed. 
When two models fit the data comparably, preferring the simpler model is not mere parsimony; it reflects the fact that the simpler model has been more strongly tested.
What we need is a framework that balances fit quality against model complexity.\footnote{Information criteria such as AIC and BIC address this problem~\cite{Akaike1974, Schwarz1978}, but rely on assumptions such as normally distributed parameters that may not hold in practice.}
Bayesian model selection provides us with a rational approach to comparing competing models, and we discuss this in Section~\ref{sec:model}.

\subsection{Parameter Estimation}

To fit a model to data, we must determine its parameter values. 
But given noisy or uncertain data, no single set of parameters is uniquely consistent with the observations---many different parameter combinations could plausibly describe the observed data. 
Rather than seeking a single ``best fit'', we want a way to characterise the full set of parameter values that are compatible with our data.
Standard fitting approaches return point estimates, perhaps with symmetric error bars that quantify spread around the best fit. 
But this representation is incomplete; it doesn't capture the full range of models consistent with the data, nor any correlations or asymmetries in how parameters relate to one another.
Bayesian posterior sampling addresses this by generating samples from the distribution of parameter values compatible with the observed data. 
This posterior distribution fully characterises our uncertainty about the model parameters, as we discuss in Section~\ref{sec:parameter_estimation}.

\subsection{Extrapolation}

Once we have fitted a temperature-dependent model, a common goal is to predict behaviour at temperatures outside the measured range; for example, estimating room-temperature conductivity from high-temperature simulations~\cite{Goldmann2025}. 
For such extrapolations to be meaningful, the uncertainties in the fitted parameters must propagate into the extrapolated predictions. 
Without these, extrapolated values cannot be meaningfully compared to other values or assessed for reliability.
Bayesian posterior sampling makes uncertainty propagation straightforward: the same parameter samples used to characterise the fit can be propagated through the model to generate a distribution of predicted values at any temperature, as we discuss in Section~\ref{sec:extrapolation}.

\section{Parameter Estimation}
\label{sec:parameter_estimation}

Although model selection conceptually precedes fitting---we need to choose a model before fitting it---in practice, we must understand how to fit a single model before we can compare models.
We therefore begin with parameter estimation.
For illustration, we assume an Arrhenius model, and demonstrate how Bayesian posterior sampling yields parameter estimates with meaningful uncertainties.
The approach is general, however, and applies to any parametric model.

Here, our goal is not to find a single ``best'' set of parameter values, but to characterise the full range of parameter values that are compatible with our data. 
For the Arrhenius model, this means finding all combinations of activation energy, $E_\mathrm{a}$, and pre-exponential factor, $A$, that could plausibly describe the observed temperature-dependent conductivity.

Given a set of measured conductivities at different temperatures, how do we determine which parameter values are compatible with these data? 
The starting point is to ask: for a given choice of $\Ea$ and $A$, how well does the model predict the observed data?
We quantify this using the likelihood: the probability of observing our data given a particular set of parameter values. This is a conditional probability, written $\pr{\bm{x}}{\bm{\theta}}$, where $\bm{x}$ denotes our data and $\bm{\theta}$ the parameters. 
If a parameter combination produces model predictions that closely match the observations (within their uncertainties), the likelihood is high; if the predictions are far from the data, the likelihood is low.

The likelihood tells us how probable our data are given some parameters. 
But this is not quite what we want. 
We want the inverse: how probable are the parameters given our data? 
This is the posterior distribution, written $\pr{\bm{\theta}}{\bm{x}}$.
Likelihood and posterior are related but not the same: $\pr{\bm{x}}{\bm{\theta}}$ is not the same as $\pr{\bm{\theta}}{\bm{x}}$. 
The relationship between them is given by Bayes' theorem:
\begin{equation}
\pr{\bm{\theta}}{\bm{x}} = \frac{\pr{\bm{x}}{\bm{\theta}} \times p(\bm{\theta})}{p(\bm{x})}.
\end{equation}
Here, $\pr{\bm{\theta}}{\bm{x}}$ is the posterior: the probability of parameters $\bm{\theta}$ given the data. 
$\pr{\bm{x}}{\bm{\theta}}$ is the likelihood. 
$p(\bm{\theta})$ is the prior: our knowledge about plausible parameter values before seeing the data. 
$p(\bm{x})$ is a normalising constant that ensures the posterior integrates to one.
Because $p(\bm{x})$ is a constant for any given dataset, the posterior is proportional to the likelihood times the prior:
\begin{equation}
\pr{\bm{\theta}}{\bm{x}} \propto \pr{\bm{x}}{\bm{\theta}} \times p(\bm{\theta}).
\end{equation}
This means we can characterise the posterior by evaluating the likelihood and prior across parameter space.

The prior distribution, $p(\bm{\theta})$, encodes what we know about plausible parameter values before examining the data. 
In many cases, we have little prior knowledge, in which case we use broad, minimally informative priors: for example, a uniform distribution over a physically reasonable range.
For the Arrhenius model, we require $\Ea > 0$ (activation energies are positive) and $A > 0$ (the pre-exponential factor is positive). 
Beyond these physical constraints, we might specify uniform priors over ranges that comfortably encompass expected values. 
Provided the data are informative, the posterior will be dominated by the likelihood, so the precise choice of prior is often not critical; but it should always be reported.

For simple models with few parameters, we could evaluate the posterior on a fine grid across parameter space.
But this becomes impractical as the number of parameters grows. 
Instead, we use sampling algorithms that explore parameter space efficiently, spending more time in regions of high posterior probability.
Markov chain Monte Carlo (MCMC) is a widely used family of sampling algorithms. 
The details of how MCMC works are beyond the scope of this tutorial, but the essential idea is that it generates a sequence of parameter samples that, collectively, represent the posterior distribution. 
From these samples, we can compute summary statistics (means, medians, credible intervals) or visualise the full distribution.

To apply this framework, we first choose both a predictive model---for example, the Arrhenius equation---and a statistical model for the uncertainties in our data.
Together, these determine the likelihood function we use in Bayes' theorem.
For a model with parameters $\bm{\theta}$, each choice of parameters predicts a conductivity value at each measured temperature; call these predictions $\bm{\mu}(\bm{\theta})$.
We also have our observed conductivity values, $\bm{x}$, which will not exactly match the predictions because of uncertainties in the data.
A standard approach is to assume that the observed values are normally distributed around the model predictions. 
This assumption is often reasonable: if errors arise from many independent sources, their combined effect will tend toward a normal distribution. More fundamentally, if all we know about our errors is their mean and variance, the normal distribution is the only distribution that requires no further assumptions.\footnote{This follows from the principle of maximum entropy~\cite{Jaynes_PhysRev1957}.}

Under this assumption, our statistical model is a multivariate normal distribution, characterised by a covariance matrix $\bm{\Sigma}$ that describes the data uncertainties.
The corresponding likelihood function is:
    \begin{equation}
        \pr{\bm{x}}{\bm{\theta}} = (2\pi)^{-\frac{k}{2}}|\bm{\Sigma}|^{-\frac{1}{2}}\exp\left\{-\frac{1}{2}\left[\bm{\mu}(\bm{\theta}) - \bm{x}\right]^\top\bm{\Sigma}^{-1}\left[\bm{\mu}(\bm{\theta}) - \bm{x}\right]\right\},
        \label{eqn:gen}
    \end{equation}
%
where $k$ is the number of observations, $\bm{x}$ is the vector of observed conductivity values, $\bm{\mu}(\bm{\theta})$ is the vector of model predictions, and $\bm{\Sigma}$ is the covariance matrix.

In practice, the procedure is as follows. 
We start with a model---here, the Arrhenius equation---and define priors for each parameter. 
We then construct the likelihood function using our measured data and their uncertainties. 
The sampler explores parameter space, evaluating the product of likelihood and prior at each point. 
Combinations that fit the data well and satisfy the prior receive high probability; those that fit poorly or lie outside the prior bounds receive low probability. 
After sufficient exploration, the collected samples represent the posterior distribution. 
From these samples, we can extract summary statistics (e.g., means, medians, credible intervals) or visualise the full distribution to understand correlations between parameters.

To illustrate this procedure, we use conductivity data from molecular dynamics simulations of a lithium-ion conductor, c-LLZO (see Methods for simulation details).
The data comprise conductivity estimates at four temperatures, each with an associated uncertainty (see Methods for analysis details).
To sample the posterior, we need to specify priors for $E_\mathrm{a}$ and $A$. 
Since we have no strong prior knowledge, we use broad uniform distributions that comfortably span the range of physically plausible values. 
With the likelihood and priors defined, we can run the sampler. 
Full details of the prior ranges and sampling procedure are given in the Appendix.

\begin{figure}
    \centering
    \includegraphics[width=\columnwidth]{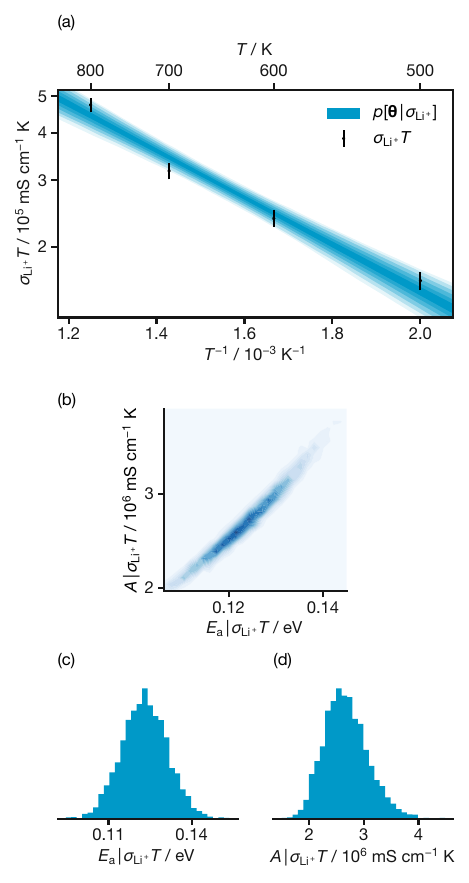}
    \vspace{-2\baselineskip}
    \caption{(a) Lithium-ion conductivity in c-LLZO (black points; error bars show \qty{95}{\percent} credible intervals). Blue curves show Arrhenius models evaluated using parameter samples drawn from the posterior; darker shading indicates where more curves overlap. (b) Joint posterior distribution showing the correlation between $\Ea$ and $A$. (c, d) Marginal posterior distributions for the activation energy and pre-exponential factor.}
    \label{fig:llzo}
    \script{llzo.py}
\end{figure}

The sampler returns a set of samples drawn from the joint posterior distribution $\pr{E_\mathrm{a}, A}{\bm{x}}$. 
Each sample is a point in parameter space: a specific ($\Ea$, $A$) combination that corresponds to a curve on the Arrhenius plot. 
\cref{fig:llzo}(a) shows the data with many such curves overlaid: the shaded bands indicate where curves cluster more or less densely, reflecting which parameter combinations are more probable given the data. 
Panel (b) shows these samples as a two-dimensional distribution, revealing that the parameters are highly correlated: higher values of $\Ea$ tend to be associated with higher values of $A$. 
Panels (c) and (d) show the marginal distributions for each parameter, obtained by integrating the samples along each axis. 
This is equivalent to integrating the joint distribution over the other parameter, asking, for example, ``what is the probability of this value of $\Ea$, regardless of $A$?''

The activation energy is approximately normally distributed, with mean and \qty{95}{\percent} credible interval of %
  $\SI{0.123\pm0.016}{\electronvolt}$\unskip\label{output/activation_energy_llzo.txt}\unskip%
. 
The pre-exponential factor, however, is not normally distributed---it follows an approximately log-normal distribution. 
For this parameter, we report the median and \qty{95}{\percent} credible interval: %
  $\left(2.644^{+0.899}_{-0.653}\right)\times10^{6}\,\si{\centi\meter^2\second^{-1}}$\unskip\label{output/preexp_llzo.txt}\unskip%
.
This illustrates one advantage of Bayesian posterior sampling: we obtain the full distribution of each parameter, rather than assuming normality.
For $A$, reporting a symmetric interval around the mean would misrepresent our uncertainty.

Arrhenius models are often fitted by taking logarithms and fitting a straight line to $\ln(\sigma T)$ versus $1/T$.
This approach is convenient, but introduces bias: if the errors in $\sigma T$ are normally distributed, transforming to log space makes them log-normally distributed, violating the assumptions of linear regression and yielding biased estimates for $\Ea$ and $A$~\cite{McCluskey2023a}. 
Our choice of a multivariate normal likelihood above also assumes normally distributed errors, so we must work with the untransformed data to satisfy this assumption.

\section{Model Selection}
\label{sec:model}

In Section~\ref{sec:parameter_estimation}, we assumed the data followed the Arrhenius model. 
But how do we know this is the right choice? 
If we simply compare how well different models fit the data, a more complex model will generally fit better~\cite{Mayer2010}, even when the additional complexity is not justified. 
We need a rational approach that balances fit quality against model complexity.
Bayesian model selection provides such a framework. 
Rather than asking ``which model fits best?'', it asks ``which model is most probable given the data?''; and in answering this question, it naturally penalises unnecessary complexity.

To illustrate model selection, we compare the Arrhenius equation with the Vogel-Tammann-Fulcher (VTF) equation~\cite{Vogel1921, Tammann1926, Fulcher1925}, an alternative model that allows the activation energy to vary with temperature. 
The VTF equation has the form:
\begin{equation}
    \sigma T = A\exp{\left[\frac{-B}{R(T-T_0)}\right]},
    \label{eqn:vtf}
\end{equation}
where $A$ is the pre-exponential factor, $B$ is related to the activation energy, and $T_0$ is the Vogel temperature. 
The Arrhenius equation is a special case of VTF with $T_0 = 0$, so the VTF model will always fit the data at least as well as Arrhenius, because it has an additional parameter to adjust. 
The question is whether that additional parameter is justified by the data.

The Bayesian approach is to ask which model is more probable given the data; that is, compute $\pr{\mathrm{model}}{\bm{x}}$, or in more compact notation, $\pr{m}{\bm{x}}$. 
Applying Bayes' theorem at the level of models rather than parameters:
\begin{equation}
\pr{m}{\bm{x}} \propto \pr{\bm{x}}{m} \times p(m).
\end{equation}
Here, $p(m)$ is our prior belief about the model before seeing the data. 
If we have no reason to prefer one model over another, we assign them equal prior probabilities. 
In that case, comparing posterior model probabilities reduces to comparing the marginal likelihood, $\pr{\bm{x}}{m}$.
The marginal likelihood is the probability of observing our data, averaged over all possible parameter values for that model. 
If we denote the parameters of model $m$ as $\bm{\theta}_m$, this is:
\begin{equation}
\pr{\bm{x}}{m} = \idotsint_{\bm{\theta}_m} \pr{\bm{x}}{\bm{\theta}_m, m} \,\, \pr{\bm{\theta}_m}{m}\,\mathrm{d}\bm{\theta}_m.
\end{equation}
The first term in the integral is the likelihood: how well parameters $\bm{\theta}_m$ predict the data. 
The second is the prior: how plausible those parameters are before seeing the data.
This integral naturally penalises complex models. 
A model with more parameters spreads its prior probability over a larger parameter space. 
Unless the data actually constrain those extra parameters, much of this space will have low likelihood, dragging down the integral. 
A simpler model spreads its prior over a smaller space; if the data fall within this space, the integral is correspondingly larger.
This is the mathematical realisation of the idea from Section~\ref{sec:problems}: a simpler model makes a more specific prediction. 
If that prediction is confirmed by the data, it provides stronger evidence than a flexible model that could have accommodated many different outcomes.

To compare two models, we compute the ratio of their marginal likelihoods:
\begin{equation}
B_{\beta\alpha} = \frac{\pr{\bm{x}}{m_\beta}}{\pr{\bm{x}}{m_\alpha}}.
\end{equation}
This ratio is called the Bayes factor. 
If $B_{\beta\alpha} > 1$, the data favour model $\beta$; if $B_{\beta\alpha} < 1$, they favour model $\alpha$. 
The magnitude indicates the strength of the evidence. 
A common heuristic, due to Kass and Raftery, is that $\ln(B_{\beta\alpha}) > 5$ (corresponding to $B_{\beta\alpha} > 150$) constitutes ``very strong'' evidence for the preferred model~\cite{Kass1995}.

The integral for the marginal likelihood can be challenging to compute, particularly for models with many parameters. 
Markov chain Monte Carlo, which we used in Section~\ref{sec:parameter_estimation} for parameter estimation, does not directly provide the marginal likelihood. 
Nested sampling is an alternative algorithm designed specifically for this purpose: it efficiently estimates the marginal likelihood while also providing posterior samples as a by-product. 
We do not detail the algorithm here, but point the reader to Sivia and Skilling for a pedagogical description~\cite{Sivia2006}.

\begin{figure}
    \centering
    \includegraphics[width=\columnwidth]{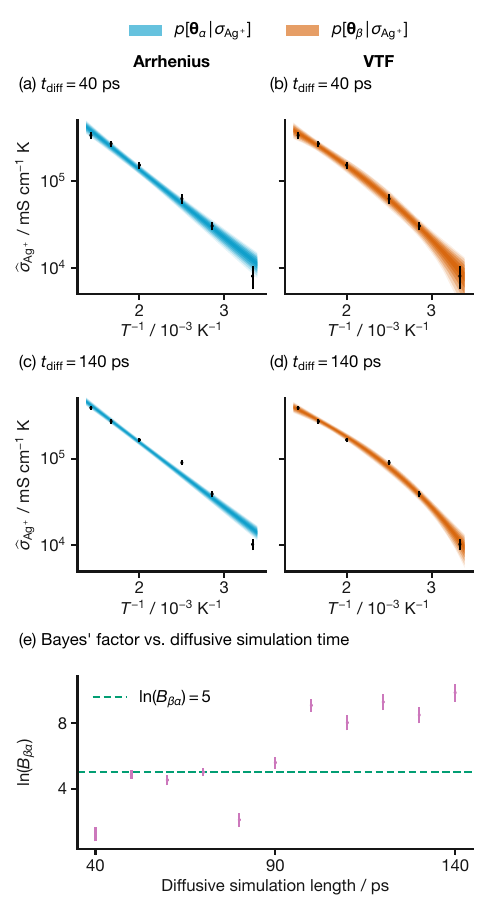}
    \vspace{-2\baselineskip}    
    \caption{Comparison of Arrhenius (a, c) and VTF (b, d) models for silver-ion conductivity in \ce{AgCrSe2}. Black points show conductivity estimates with \qty{95}{\percent} credible intervals. Coloured shading shows models evaluated at parameter values sampled from each posterior. Panels (a, b) use data from short simulations (40 ps fitting windows; see Methods); panels (c, d) use longer simulations (140 ps fitting windows). (e) Bayes factor comparing VTF to Arrhenius as a function of the fitting window width. The dashed line indicates $\ln(B_{\beta\alpha}) = 5$, the threshold for ``very strong'' evidence. Error bars represent a \qty{95}{\percent} confidence interval from 10 nested sampling repeats and reflect stochastic variability in the algorithm, not data uncertainty.}
    \label{fig:agcrse2}
    \script{agcrse2.py}
\end{figure}

As an example of this approach, we consider silver-ion conductivity data from molecular dynamics simulations of \ce{AgCrSe2}~\cite{Wang2021}.
\cref{fig:agcrse2}(a) and (b) show ionic conductivities obtained from relatively short simulation trajectories (see Methods for details), fitted using the Arrhenius and VTF models, respectively.
Both models fit the data reasonably well, so how do we choose between them?
The Bayes factor comparing the VTF model ($\beta$) to the Arrhenius model ($\alpha$), assuming equal prior probabilities for each model, is $\ln{(B_{\beta\alpha})}=\;$%
  $\num{1.2\pm0.4}$\unskip\label{output/bayes_40.txt}\unskip%
:
this indicates weak evidence for the VTF model, but is well below the threshold for strong evidence.
With this amount of data, we therefore have no compelling evidential basis for preferring the more complex VTF model.

Panels (c) and (d) show the same comparison using ionic conductivities from longer simulation trajectories.
The conductivity data appear similar to those in panels (a) and (b), but the uncertainties are much smaller, and the deviation from linearity is more clearly resolved.
The Bayes factor increases to $\ln(B_{\beta\alpha}) =\,%
  $\num{9.9\pm0.5}$\unskip\label{output/bayes_140.txt}\unskip%
$, well above the threshold of \num{5}, providing strong evidence for the more complex VTF model over the simpler Arrhenius alternative.

Panel (e) shows how the Bayes factor evolves with increasing simulation trajectory lengths.
As the conductivity data become more precise, the evidence for non-Arrhenius behaviour grows.
If the system exhibited true Arrhenius behaviour, longer simulations would not increase the Bayes factor, and we would continue to prefer the simpler model.

This example illustrates two points. 
First, Bayesian model selection tells us whether a more complex model is supported by the data or if we should prefer a simpler alternative.
Second, it tells us whether our data are sufficient to answer the question we are asking. 
With short simulation trajectories, we cannot confidently classify this behaviour as non-Arrhenius; with longer trajectories, we can.
More generally, this approach can be applied to any set of competing models; we need only compute the marginal likelihood for each.

\section{Extrapolation}
\label{sec:extrapolation}
In Sections~\ref{sec:parameter_estimation} and \ref{sec:model}, we fitted models to data within the measured temperature range. 
Often, however, we want to predict behaviour at temperatures outside this range: for example, estimating room-temperature conductivity from high-temperature simulations. 
For such extrapolations to be meaningful, the uncertainties in the fitted parameters must propagate into the predicted values---a single extrapolated number without an associated uncertainty cannot be compared to other values or assessed for reliability.

Bayesian posterior sampling makes uncertainty propagation straightforward. 
In Section~\ref{sec:parameter_estimation}, we obtained samples from the joint posterior distribution of parameters. 
Each sample represents a plausible ($\Ea$, $A$) combination given the data. 
To extrapolate, we simply evaluate the model at the new temperature for each sample. 
The resulting collection of conductivity values represents the distribution of model predictions at that temperature; i.e., the range of conductivities compatible with our original data~\footnote{This extrapolated distribution captures parameter uncertainty but not estimated measurement noise. For extrapolation far from the measured range, parameter uncertainty typically dominates, so this distinction is rarely important in practice.}.

Returning to the LLZO example from Section~\ref{sec:parameter_estimation}, we now extrapolate from the measured temperature range (\qtyrange{500}{800}{\kelvin}) to \qty{300}{\kelvin}, well outside the original data.
For each posterior sample, we calculate $\sigma(\qty{300}{\kelvin})$ using the Arrhenius equation.
These predicted values form a distribution of conductivities at room temperature.

\begin{figure}
    \centering
    \includegraphics[width=\columnwidth]{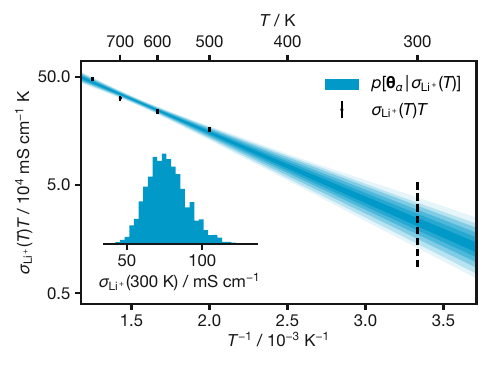}
    \vspace{-2\baselineskip}
    \caption{Extrapolation of the LLZO conductivity model to \qty{300}{\kelvin} (vertical dashed line). Blue shading shows Arrhenius models evaluated at parameter values sampled from the posterior (same data as \cref{fig:llzo}). The shaded region widens at lower temperatures, reflecting increased uncertainty in the extrapolated predictions. Inset: distribution of predicted conductivity at \qty{300}{\kelvin}.}
    \label{fig:extrapolate}
    \script{extrapolate.py}
\end{figure}

Fig.~\ref{fig:extrapolate} shows the result. 
The shaded bands in the main panel show the distribution of extrapolated conductivities at each temperature, and widen substantially as we move further from the measured range. 
The inset shows the distribution of predicted conductivity at \qty{300}{\kelvin}. 
This distribution is broad and asymmetric, with a median and \qty{95}{\percent} credible interval of %
  $\big(75.8^{+30.6}_{-22.5}\big)\,\si{\milli\siemens\centi\meter^{-1}}$\unskip\label{output/d_300.txt}\unskip%
.
The breadth of this distribution illustrates why uncertainty propagation matters. 
A point estimate at \qty{300}{\kelvin} would suggest a precise prediction; the full distribution reveals that the extrapolated conductivity spans nearly an order of magnitude. 
This uncertainty is not a failure of the method; it accurately reflects how much the data can tell us about behaviour at temperatures we did not measure.

A caveat: for extrapolation to be meaningful, the fitted model must remain valid at the new temperature.
Bayesian methods, however, cannot verify this assumption; if the transport mechanism changes---for example, if a different diffusion pathway becomes dominant at lower temperatures---the extrapolation will be misleading regardless of how carefully we propagate uncertainty.

\section{Conclusions}
\label{sec:conc}

This tutorial has introduced Bayesian methods for analysing temperature-dependent transport data, addressing three common challenges: estimating model parameters with meaningful uncertainties, selecting between competing models, and extrapolating predictions to unmeasured temperatures.
For parameter estimation, Bayesian posterior sampling returns the full distribution of parameter values compatible with the data, yielding meaningful uncertainties and capturing asymmetries and correlations that point estimates with symmetric error bars would miss.
The approach requires only a model equation and appropriate priors, and applies to any parametric model.
For model selection, Bayes factors provide a rational basis for choosing between models of different complexity, based on the relative evidence provided by the data.
For extrapolation, the same posterior samples enable straightforward uncertainty propagation, revealing how much (or how little) the data constrain predictions outside the measured range.
The examples in this tutorial have focused on transport data from molecular dynamics simulations, but the same framework applies to any temperature-dependent measurements with associated uncertainties, such as reaction rates or mechanical properties.
The methods described here are implemented in the open-source Python package \textsc{kinisi}~\cite{McCluskey2024}, which we hope will make Bayesian analysis of transport data more broadly accessible.

\section*{Author Contributions}

CRediT author statement: 
A.R.M.: Conceptualization, Formal Analysis, Investigation, Methodology, Software, Visualization, Writing - original draft. 
S.W.C.: Methodology, Resources, Writing - review and editing. 
B.J.M.: Conceptualization, Methodology, Software, Writing - review and editing.

\section*{Acknowledgements}

We thank the authors of the \ce{AgCrSe2} work for providing simulation trajectories.
This work used the Isambard 2 UK National Tier-2 HPC Service (http://gw4.ac.uk/isambard/) operated by GW4 and the UK Met Office and funded by EPSRC (EP/T022078/1). 
The authors acknowledge the University of Bath's Research Computing Group for their support in running the LLZO molecular dynamics simulations. 
S.W.C. and B.J.M. acknowledge the support of the Faraday Institution through the CATMAT Project (Grant FIRG016). 
B.J.M. acknowledges support from the Royal Society (UF130329 and URF\textbackslash R\textbackslash 191006).
This work has benefitted from the input of all members of the \textsc{kinisi} development team.

\section*{Methods}
\label{sec:meth}

All conductivity data used in this work were derived from molecular dynamics simulations.
At each temperature, self-diffusion coefficients were estimated from the long-time slope of the mean-squared displacement (MSD) using approximate Bayesian regression~\cite{McCluskey2024a}, considering only the diffusive regime identified by visual inspection of log-log MSD plots. The range of lag times used for this regression is the fitting window.
These were converted to conductivity estimates via the Nernst-Einstein relation, assuming a Haven ratio $H_\mathrm{R} = 1$.\footnote{A temperature-independent Haven ratio would uniformly scale the conductivity values without affecting the temperature dependence; a temperature-dependent ratio would introduce additional curvature, but the analysis procedure would be unchanged.}
All analysis was performed with \textsc{kinisi}-2.0.3~\cite{McCluskey2024}, using \textsc{MDAnalysis}~\cite{MichaudAgrawal2011,Gowers2016} for trajectory parsing; scripts are available in the Electronic Supplementary Information~\cite{McCluskey2025}.

For \ce{Li7La3Zr2O12} (LLZO), classical molecular dynamics simulations were run using the \textsc{metalwalls} code~\cite{MarinLaflche2020}.
These simulations used the DIPPIM polarisable ion force field, as parameterised by Burbano \emph{et al.}~\cite{Burbano2016}.
The cubic phase of LLZO was simulated in the \emph{NVT} ensemble at temperatures of \qtylist{500;600;700;800}{\kelvin}.
Simulations were run for \qty{50}{\pico\second} using a \qty{0.5}{\femto\second} timestep and a Nosé-Hoover thermostat with a relaxation time of \qty{121}{\femto\second}~\cite{Nos1984, Hoover1985, Martyna1992}. 
The simulations used $2 \times 2 \times 2$ supercells with \num{1536} atoms, following the protocol in Ref.~\onlinecite{Burbano2016}.
The simulation data are available under a CC BY 4.0 license on Zenodo~\cite{Coles2025}.
Mean-squared displacements of the lithium ions were computed at lag times from \qty{0.5}{\femto\second} to \qty{50}{\pico\second} in intervals of \qty{2.5}{\femto\second}, with the start of the fitting window at \qty{10}{\pico\second}.

The LLZO conductivity data were fitted to the Arrhenius equation using Markov chain Monte Carlo sampling.
Parameter priors were uniform: $p(E_\mathrm{a}) \sim \mathcal{U}({0},\qty{0.5}{\electronvolt})$ and $p(A) \sim \mathcal{U}(\num{e5},\qty{e7}{\centi\meter\squared\per\second})$.
The posterior distribution was sampled using \num{32} walkers, generating \num{10000} samples after a burn-in period of \num{500} samples.
Samples were thinned by a factor of \num{10} to reduce autocorrelation. 
For extrapolation, posterior samples were used to estimate conductivity at $T=\qty{300}{\kelvin}$.

For \ce{AgCrSe2}, simulation trajectories were provided by the authors of Ref.~\onlinecite{Wang2021}.
These simulations covered temperatures of \SIlist{300;350;400;500;600;700}{\kelvin}; full details are given in the original publication. 
The trajectories are available on Zenodo~\cite{Wang2025}.

To investigate how simulation length affects the Bayes factor, we truncated the simulation trajectories to different lengths, setting $t_\mathrm{sim} = \Delta t_\mathrm{min}(T) + \Delta t_\mathrm{diff}$, where $\Delta t_\mathrm{min}(T)$ is the temperature-dependent onset of the diffusive regime (\cref{tab:lengths}).
This gives a fitting window of width $\Delta t_\mathrm{diff}$, which we varied from \qty{40}{\pico\second} to \qty{140}{\pico\second} (\cref{fig:agcrse2}e).
At each temperature, MSDs were computed at lag times from $\Delta t_\mathrm{min}$ to $t_\mathrm{sim}$---i.e., we always fit to the end of the available data---in increments of $\Delta(\Delta t)$ (\cref{tab:lengths}).

Bayesian model selection was performed using nested sampling~\cite{Skilling2004} as implemented in the \textsc{dynesty} Python package~\cite{Speagle2020}, with \num{500} live points and a stopping criterion of \num{0.01}.
For the Arrhenius model, the parameter priors were: $p(E_\mathrm{a}) \sim \mathcal{U}(0, \qty{1}{\electronvolt})$ and $p(A) \sim \mathcal{U}(\num{e4}, \qty{e8}{\centi\meter\squared\per\second})$.
For the VTF model: $p(B) \sim \mathcal{U}(0, \qty{1}{\electronvolt})$, $p(A) \sim \mathcal{U}(\num{e4}, \qty{e8}{\centi\meter\squared\per\second})$, and $p(T_0) \sim \mathcal{U}(0, \qty{300}{\kelvin})$.

\begin{table}
    \centering
    \caption{MSD fitting parameters for \ce{AgCrSe2} at each simulation temperature. $\Delta t_\mathrm{min}$ is the minimum lag time used for fitting, corresponding to the onset of the diffusive regime. $\Delta(\Delta t)$ is the spacing between successive lag times in the MSD regression.}
    \label{tab:lengths}
    \setlength{\tabcolsep}{1em}
    \begin{tabular}{ccc}
        \toprule
        $T$ / K & $\Delta t_\mathrm{min}$ / ps & $\Delta(\Delta t)$ / ps \\
        \midrule
        300 & 70.0 & %
  \num{3.00}\unskip\label{output/dt300.txt}\unskip%
 \\
        350 & 50.0 & %
  \num{1.25}\unskip\label{output/dt350.txt}\unskip%
 \\
        400 & 30.0 &%
  \num{0.60}\unskip\label{output/dt400.txt}\unskip%
 \\
        500 & 25.0 & %
  \num{0.30}\unskip\label{output/dt500.txt}\unskip%
 \\
        600 & 22.5 & %
  \num{0.25}\unskip\label{output/dt600.txt}\unskip%
 \\
        700 & 20.0 & %
  \num{0.20}\unskip\label{output/dt700.txt}\unskip%
 \\
        \bottomrule
    \end{tabular}
\end{table}

\bibliographystyle{naturemag}
\bibliography{bib}

\let\addcontentsline\oldaddcontentsline

\onecolumngrid
\clearpage 
\twocolumngrid

\appendix
\renewcommand\thesection{SI.\Roman{section}}
\counterwithout{figure}{section}
\renewcommand\thefigure{SI.\arabic{figure}}
\setcounter{figure}{0}
\counterwithout{equation}{section}
\renewcommand\theequation{SI.\arabic{equation}}
\setcounter{equation}{0}
\counterwithout{table}{section}
\renewcommand\thetable{SI.\arabic{table}}
\setcounter{table}{0}
\pagenumbering{arabic} 
\renewcommand\thepage{SI.\arabic{page}}
\addtocontents{toc}{\protect\setcounter{tocdepth}{0}}

\title{Supplemental Material for ``\papertitle''}
\maketitle

This document presents supplementary material for the manuscript ``\papertitle''.
It contains the following sections:
\renewcommand{\labelitemi}{}
\begin{itemize}
    \item \ref{sec:llzo_plots}. Plots showing the mean-squared displacement and resulting linear model posterior distributions as a function of temperature for the LLZO systems.
    \item \ref{sec:plots}. Plots showing the mean-squared displacement as a function of temperature, posterior distributions for $\sigma$, and the resulting Arrhenius, VTF models and distributions for $\Ea$ for a series of different amounts of diffusive simulation time for \ce{AgCrSe2}. 
\end{itemize}
A repository containing the analysis and plotting code used to generate all results and figures in the main manuscript and this supplemental material document is available at \url{www.github.com/arm61/model-arrhenius}~\cite{McCluskey2025}, under MIT (code) and CC BY-SA 4.0 (figures and text) licenses.
This repository includes a fully reproducible \textsc{showyourwork} workflow, which allows complete reproduction of the analysis, plotting of figures and compilation of the manuscript.

\twocolumngrid

\section{Mean-squared displacement of LLZO as a function of temperature}
\label{sec:llzo_plots}

\begin{figure*}
    \vspace{-1.5\baselineskip}
    \centering
    \includegraphics[width=\textwidth]{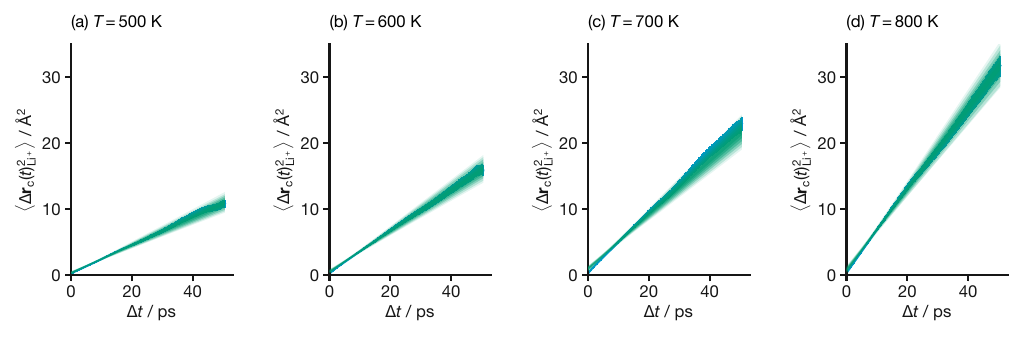}
    \vspace{-2\baselineskip}
    \script{llzo_sim.py}
    \caption{The mean squared displacement data and associated distribution of linear models at temperatures of (a)~\SI{500}{\kelvin}, (b)~\SI{600}{\kelvin}, (c)~\SI{700}{\kelvin}, and (d)~\SI{800}{\kelvin} with \SI{50}{\pico\second} of LLZO simulation.}
    \label{fig:llzo_sim}
    \vspace{-1.5\baselineskip}
\end{figure*}

\cref{fig:llzo_sim} presents the mean-squared displacement data and resulting posterior distribution of linear models for a range of temperatures for LLZO. 
The marginal posterior distribution for $\D$ is then used to find $\sigma$, which is included in the LLZO analysis in the main text (\cref{fig:llzo} and \cref{fig:extrapolate}).

\section{Temperature dependence as a function of diffusive simulation time for \ce{AgCrSe2}}
\label{sec:plots}

\begin{figure*}
    \vspace{-1.5\baselineskip}
    \centering
    \includegraphics[width=\textwidth]{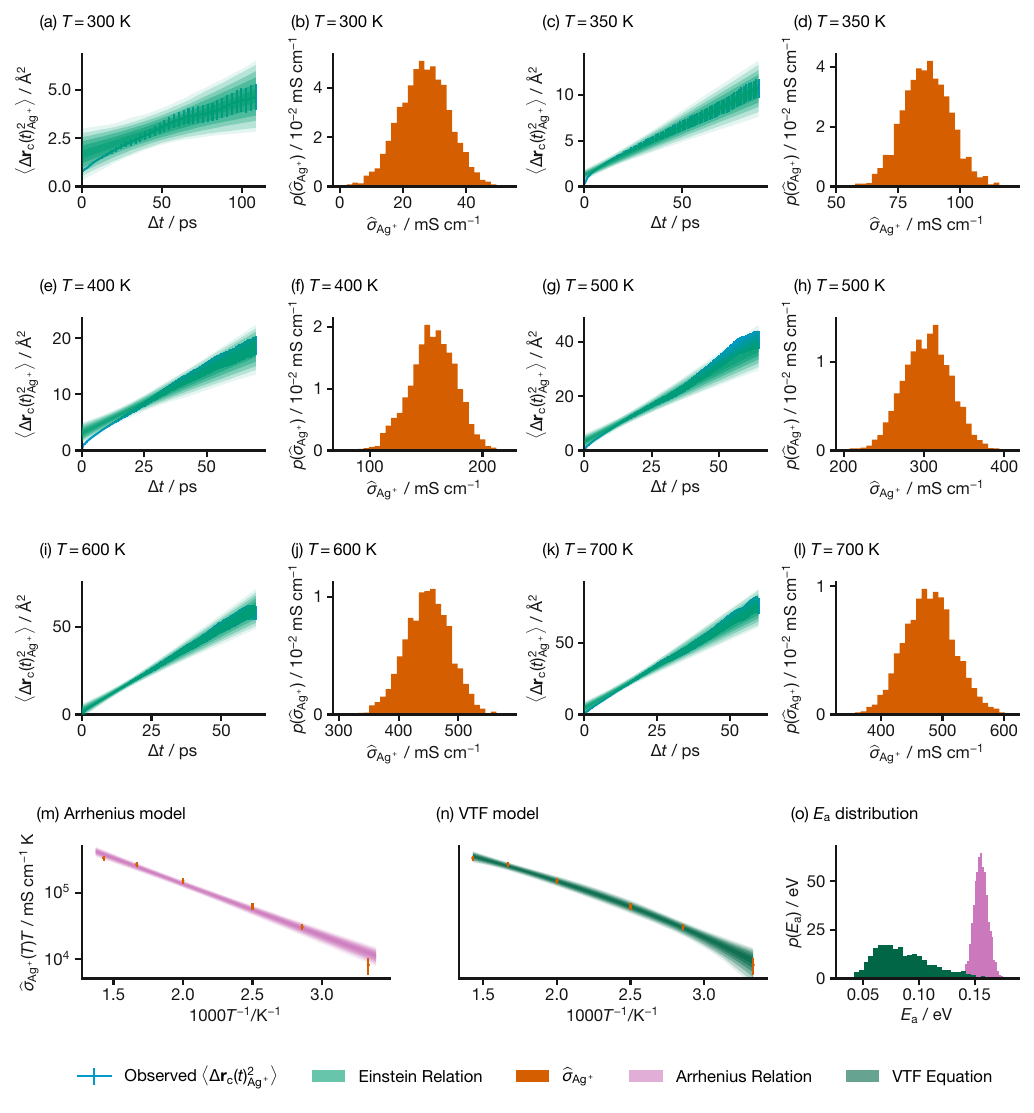}
    \vspace{-2\baselineskip}
    \script{simulation.py}
    \caption{The mean squared displacement data and associated $\sigma$ distributions at temperatures of \SIlist{300;350;400;500;600;700}{\kelvin} with \SI{40}{\pico\second} of diffusive simulation (a-l), the appropriate Arrhenius (m) and VTF model (n) plots and the resulting distributions of $\Ea$ from each modelling approach (o).}
    \label{fig:agcrse2_40}
    \vspace{-1.5\baselineskip}
\end{figure*}

\begin{figure*}
    \vspace{-1.5\baselineskip}
    \centering
    \includegraphics[width=\textwidth]{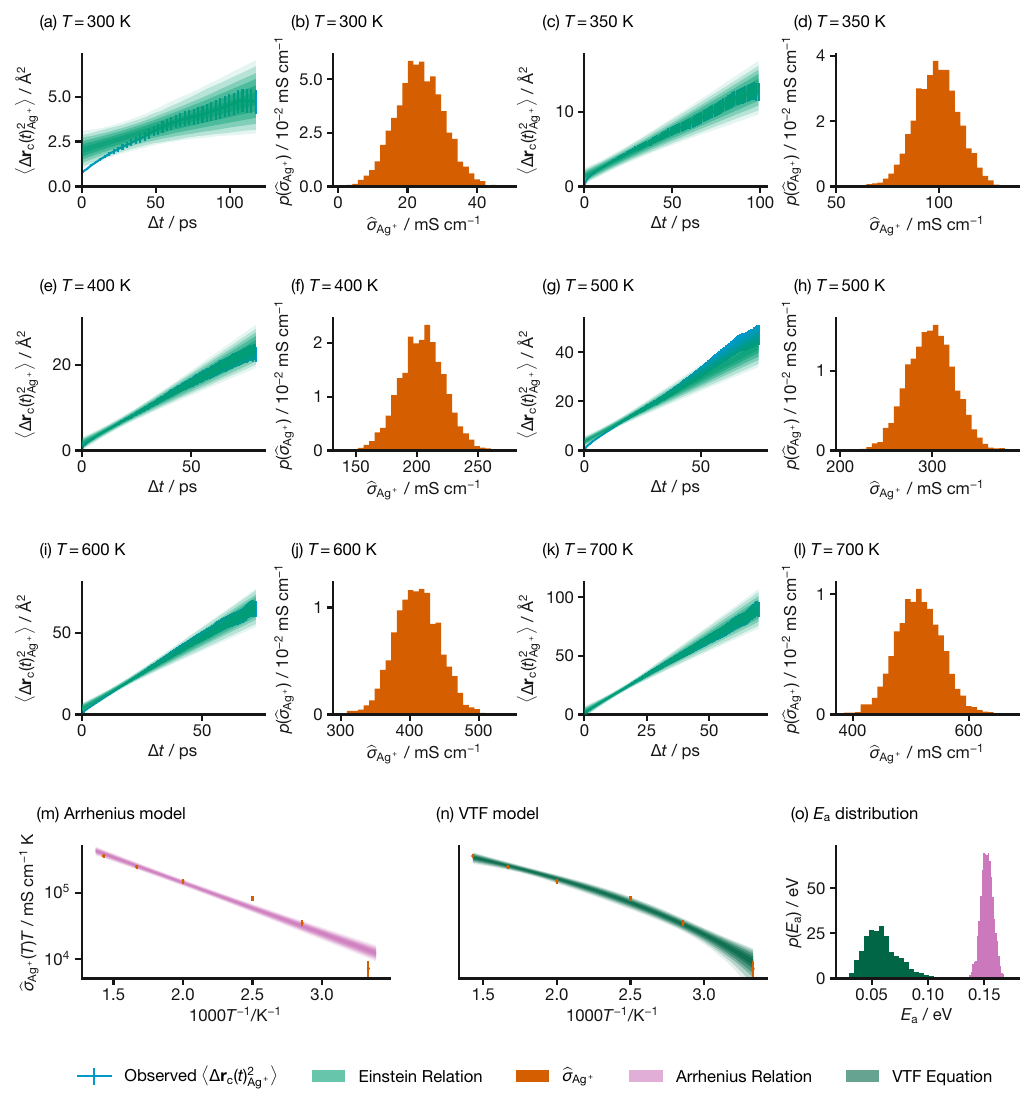}
    \vspace{-2\baselineskip}
    \script{simulation.py}
    \caption{The mean squared displacement data and associated $\sigma$ distributions at temperatures of \SIlist{300;350;400;500;600;700}{\kelvin} with \SI{50}{\pico\second} of diffusive simulation (a-l), the appropriate Arrhenius (m) and VTF model (n) plots and the resulting distributions of $\Ea$ from each modelling approach (o).}
    \label{fig:agcrse2_50}
    \vspace{-1.5\baselineskip}
\end{figure*}

\begin{figure*}
    \vspace{-1.5\baselineskip}
    \centering
    \includegraphics[width=\textwidth]{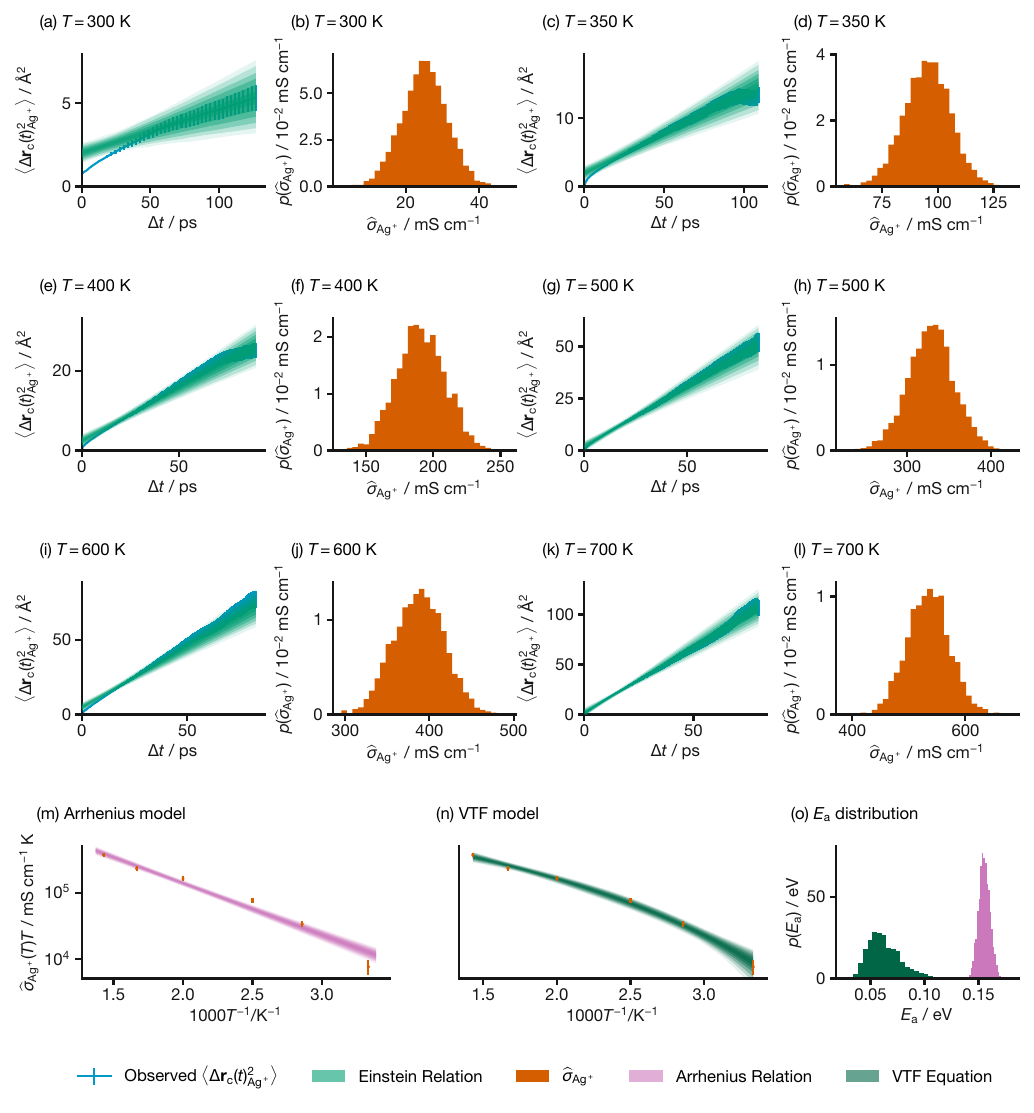}
    \vspace{-2\baselineskip}
    \script{simulation.py}
    \caption{The mean squared displacement data and associated $\sigma$ distributions at temperatures of \SIlist{300;350;400;500;600;700}{\kelvin} with \SI{60}{\pico\second} of diffusive simulation (a-l), the appropriate Arrhenius (m) and VTF model (n) plots and the resulting distributions of $\Ea$ from each modelling approach (o).}
    \label{fig:agcrse2_60}
    \vspace{-1.5\baselineskip}
\end{figure*}

\begin{figure*}
    \vspace{-1.5\baselineskip}
    \centering
    \includegraphics[width=\textwidth]{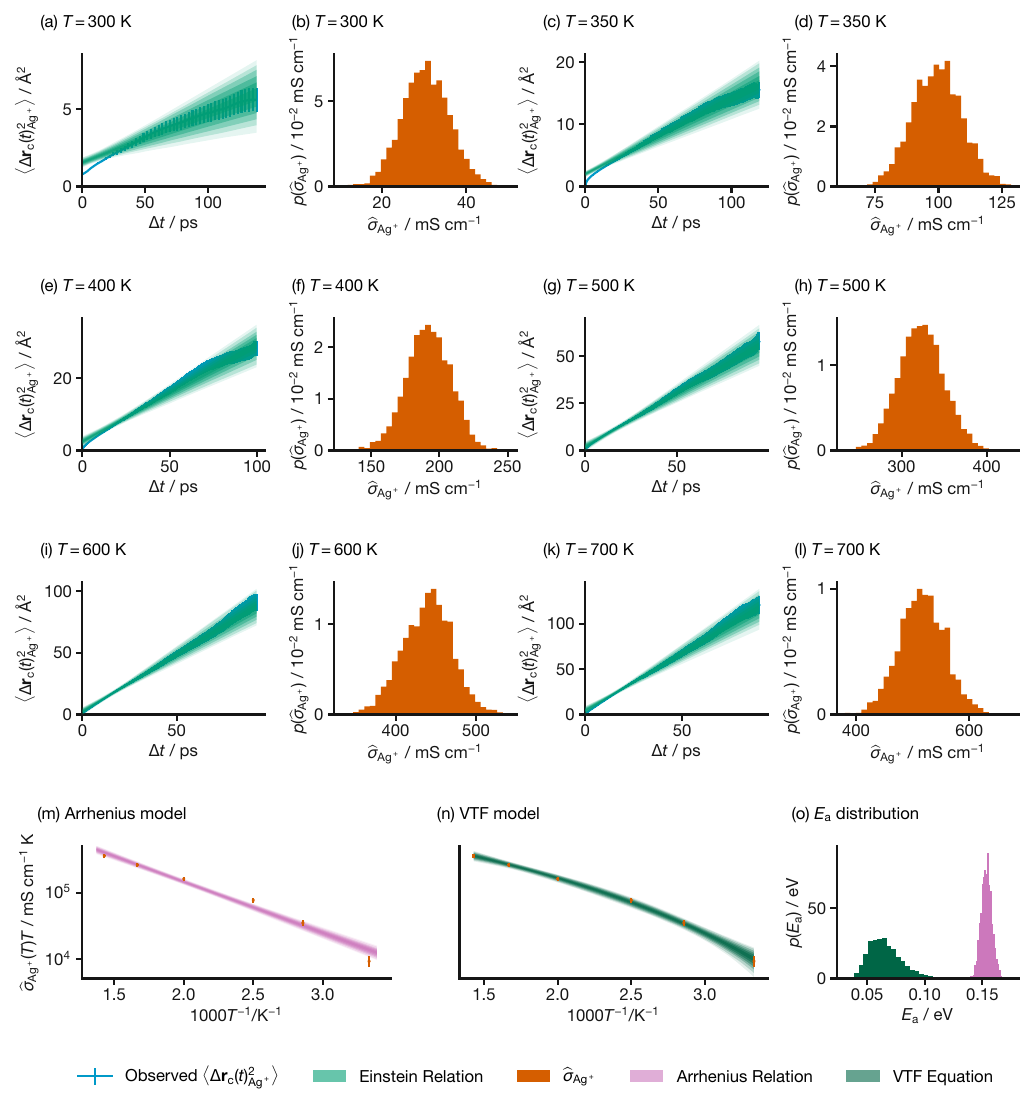}
    \vspace{-2\baselineskip}
    \script{simulation.py}
    \caption{The mean squared displacement data and associated $\sigma$ distributions at temperatures of \SIlist{300;350;400;500;600;700}{\kelvin} with \SI{70}{\pico\second} of diffusive simulation (a-l), the appropriate Arrhenius (m) and VTF model (n) plots and the resulting distributions of $\Ea$ from each modelling approach (o).}
    \label{fig:agcrse2_70}
    \vspace{-1.5\baselineskip}
\end{figure*}

\begin{figure*}
    \vspace{-1.5\baselineskip}
    \centering
    \includegraphics[width=\textwidth]{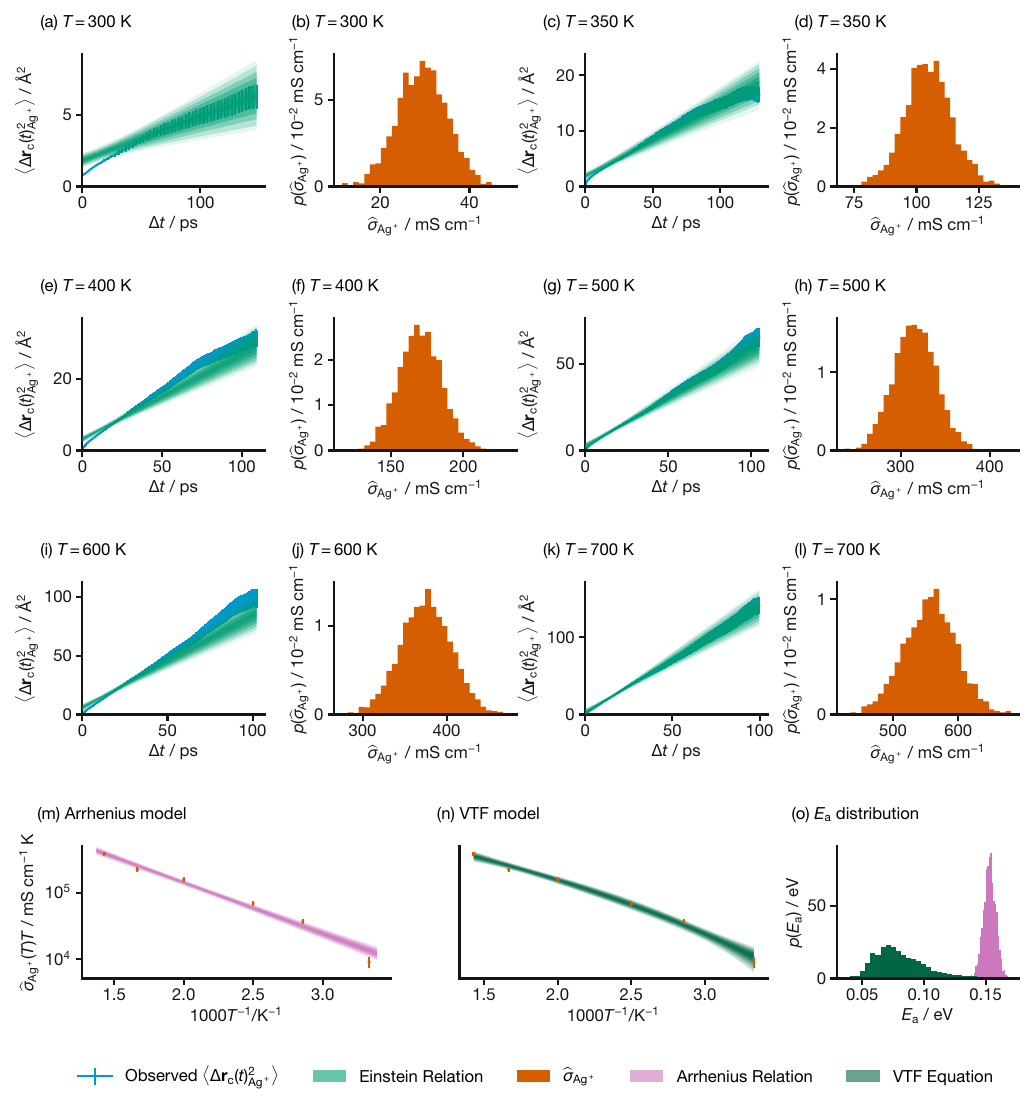}
    \vspace{-2\baselineskip}
    \script{simulation.py}
    \caption{The mean squared displacement data and associated $\sigma$ distributions at temperatures of \SIlist{300;350;400;500;600;700}{\kelvin} with \SI{80}{\pico\second} of diffusive simulation (a-l), the appropriate Arrhenius (m) and VTF model (n) plots and the resulting distributions of $\Ea$ from each modelling approach (o).}
    \label{fig:agcrse2_80}
    \vspace{-1.5\baselineskip}
\end{figure*}

\begin{figure*}
    \vspace{-1.5\baselineskip}
    \centering
    \includegraphics[width=\textwidth]{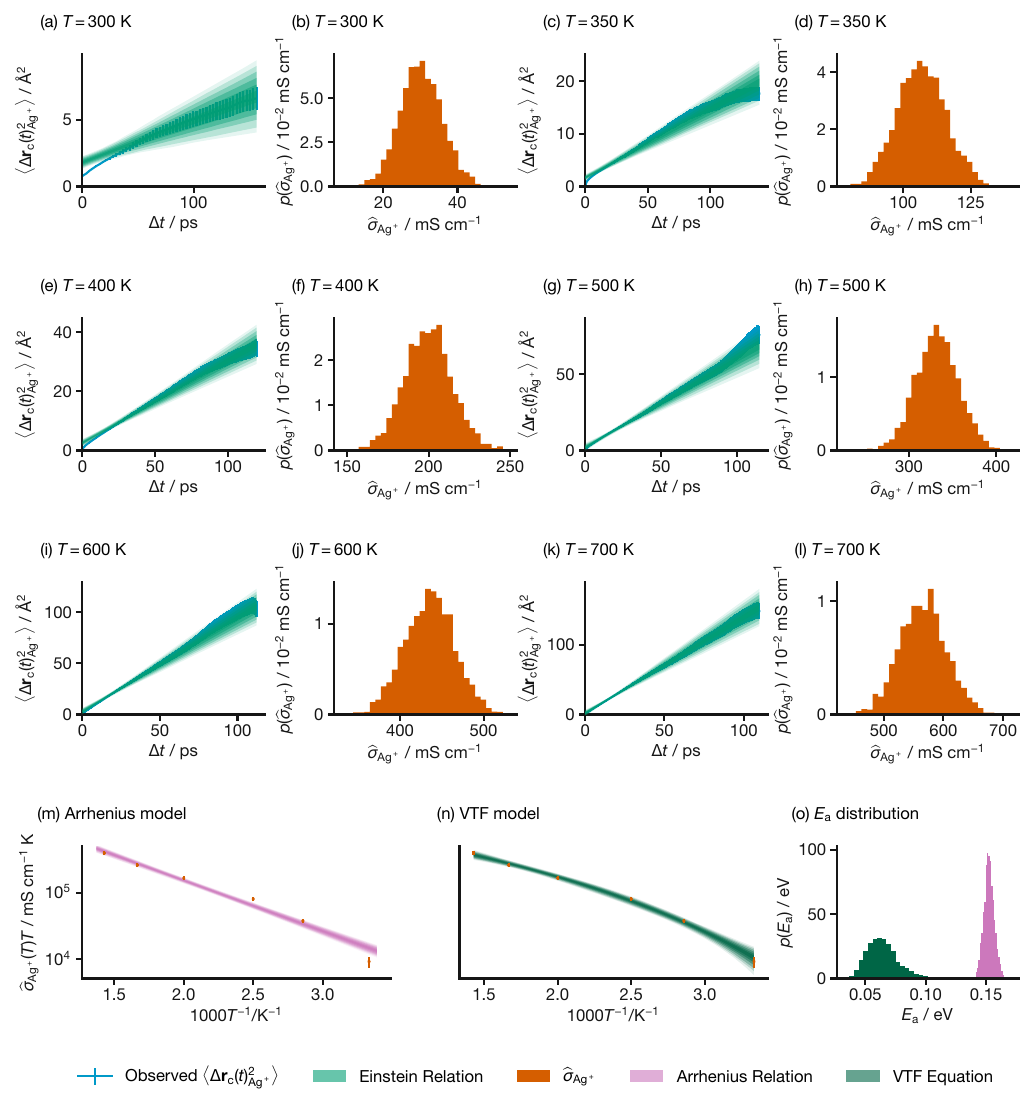}
    \vspace{-2\baselineskip}
    \script{simulation.py}
    \caption{The mean squared displacement data and associated $\sigma$ distributions at temperatures of \SIlist{300;350;400;500;600;700}{\kelvin} with \SI{90}{\pico\second} of diffusive simulation (a-l), the appropriate Arrhenius (m) and VTF model (n) plots and the resulting distributions of $\Ea$ from each modelling approach (o).}
    \label{fig:agcrse2_90}
    \vspace{-1.5\baselineskip}
\end{figure*}

\begin{figure*}
    \vspace{-1.5\baselineskip}
    \centering
    \includegraphics[width=\textwidth]{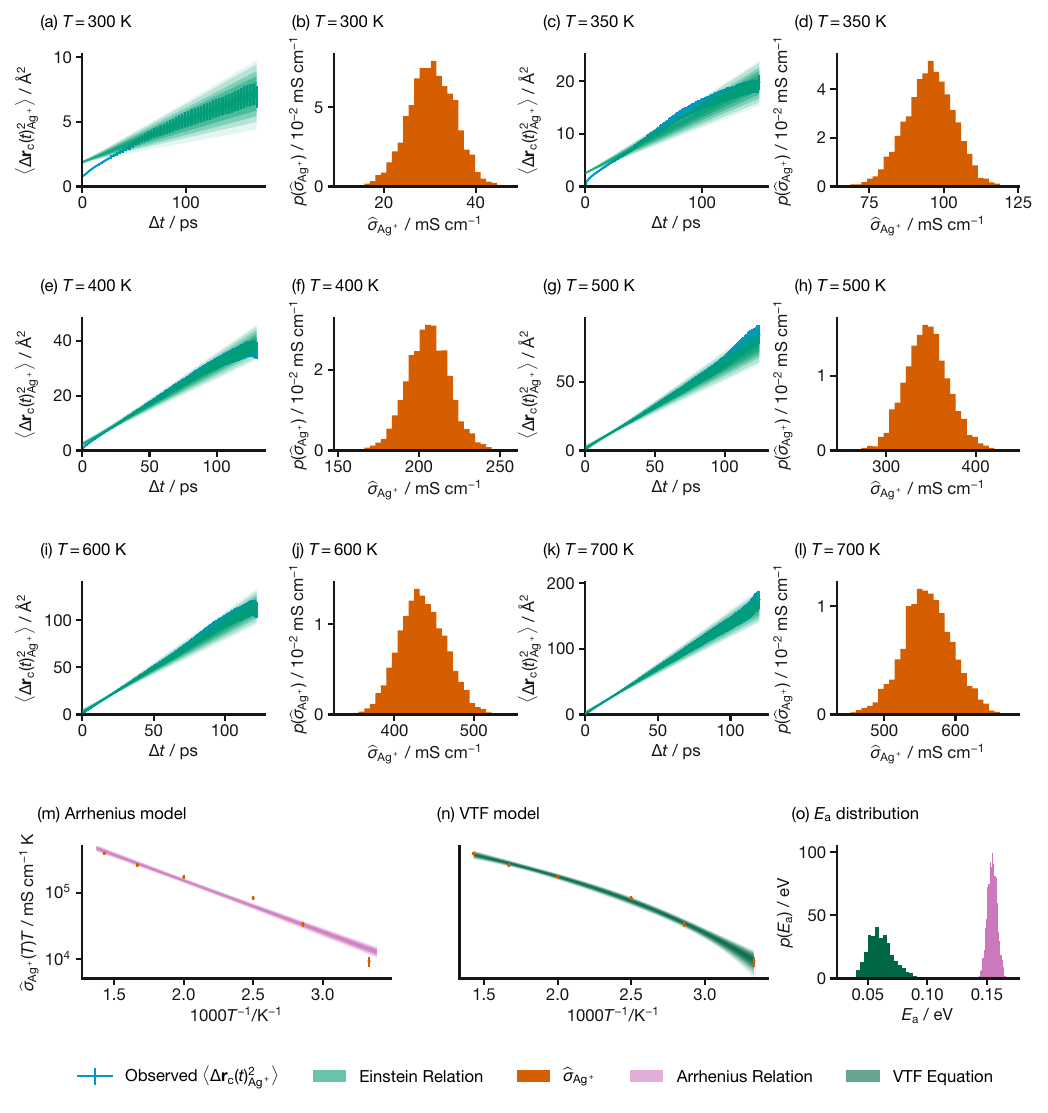}
    \vspace{-2\baselineskip}
    \script{simulation.py}
    \caption{The mean squared displacement data and associated $\sigma$ distributions at temperatures of \SIlist{300;350;400;500;600;700}{\kelvin} with \SI{100}{\pico\second} of diffusive simulation (a-l), the appropriate Arrhenius (m) and VTF model (n) plots and the resulting distributions of $\Ea$ from each modelling approach (o).}
    \label{fig:agcrse2_100}
    \vspace{-1.5\baselineskip}
\end{figure*}

\begin{figure*}
    \vspace{-1.5\baselineskip}
    \centering
    \includegraphics[width=\textwidth]{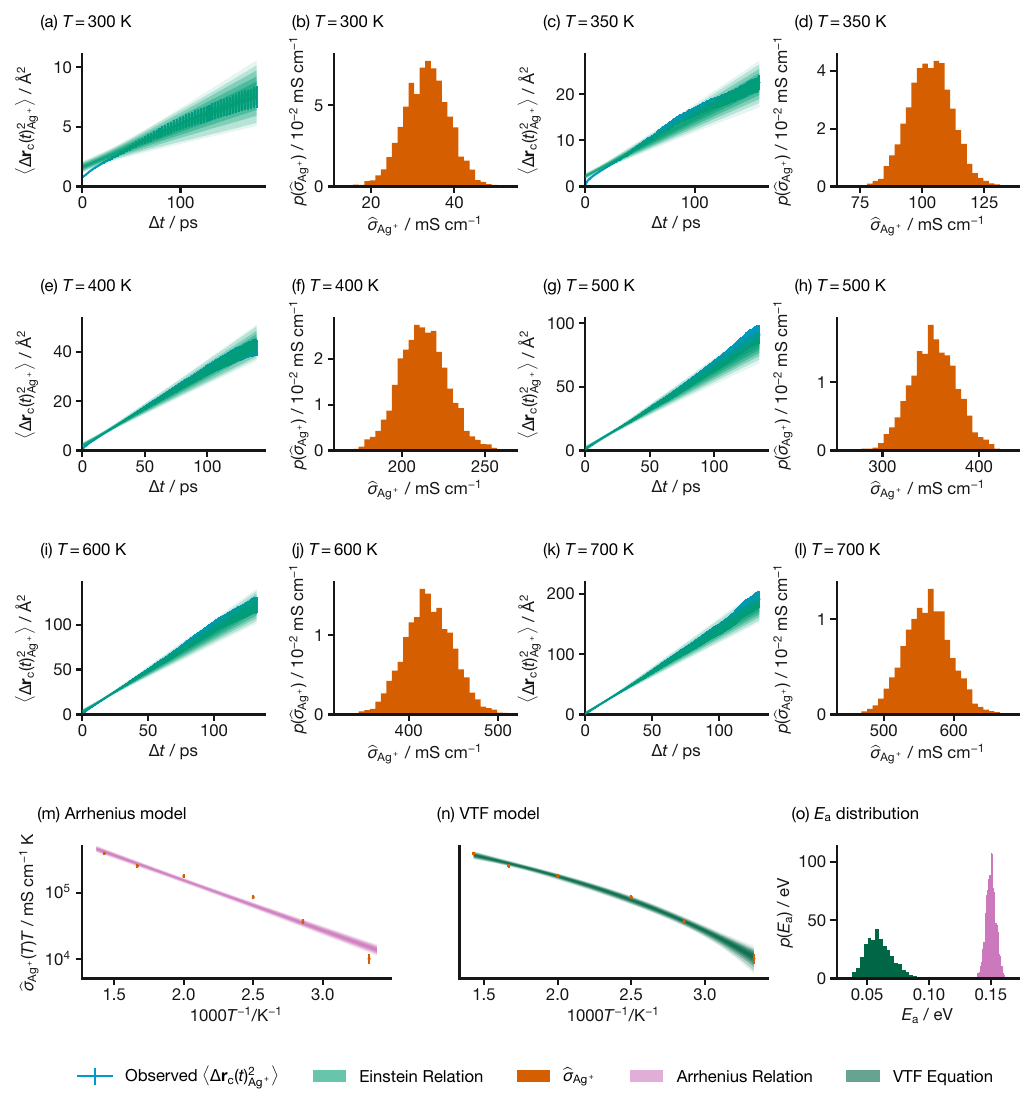}
    \vspace{-2\baselineskip}
    \script{simulation.py}
    \caption{The mean squared displacement data and associated $\sigma$ distributions at temperatures of \SIlist{300;350;400;500;600;700}{\kelvin} with \SI{110}{\pico\second} of diffusive simulation (a-l), the appropriate Arrhenius (m) and VTF model (n) plots and the resulting distributions of $\Ea$ from each modelling approach (o).}
    \label{fig:agcrse2_110}
    \vspace{-1.5\baselineskip}
\end{figure*}

\begin{figure*}
    \vspace{-1.5\baselineskip}
    \centering
    \includegraphics[width=\textwidth]{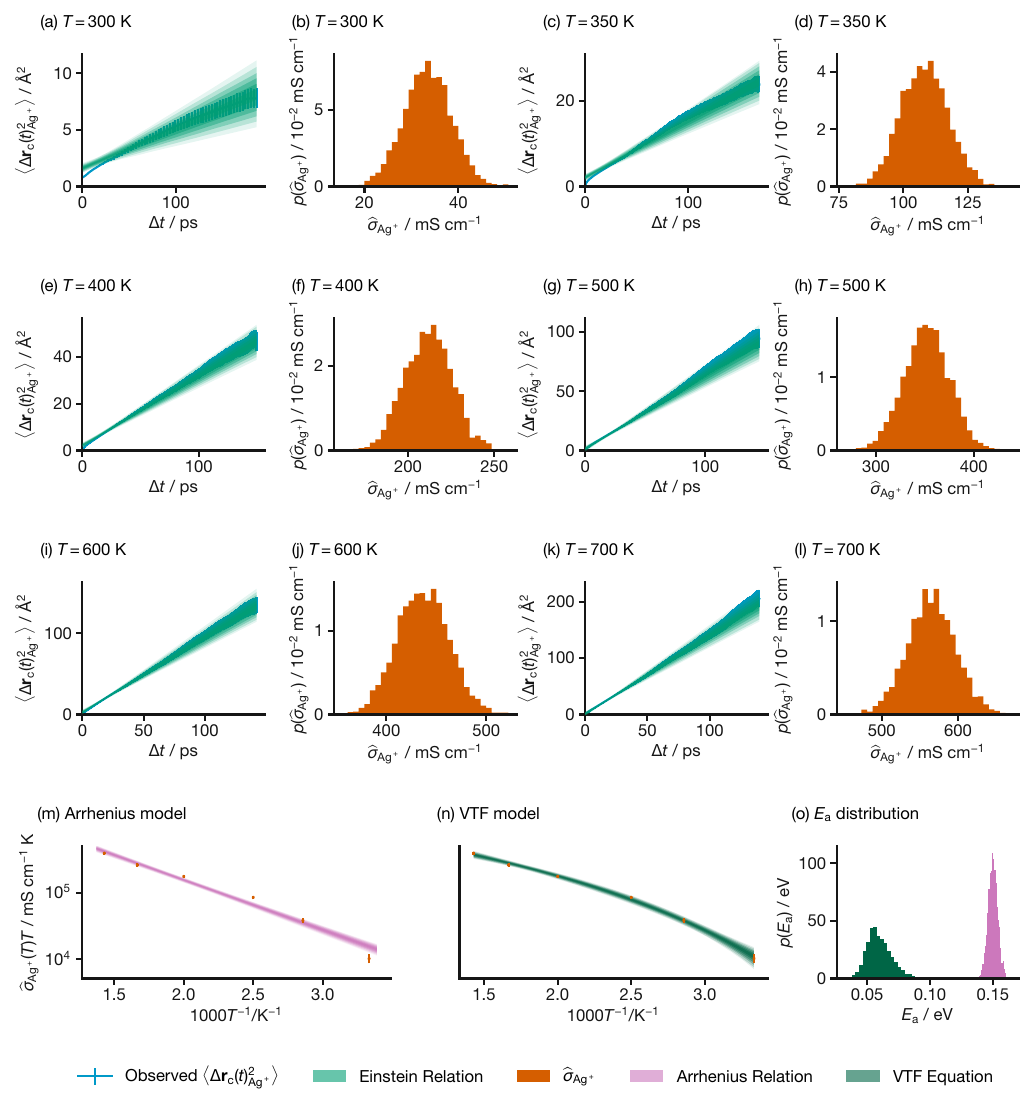}
    \vspace{-2\baselineskip}
    \script{simulation.py}
    \caption{The mean squared displacement data and associated $\sigma$ distributions at temperatures of \SIlist{300;350;400;500;600;700}{\kelvin} with \SI{120}{\pico\second} of diffusive simulation (a-l), the appropriate Arrhenius (m) and VTF model (n) plots and the resulting distributions of $\Ea$ from each modelling approach (o).}
    \label{fig:agcrse2_120}
    \vspace{-1.5\baselineskip}
\end{figure*}

\begin{figure*}
    \vspace{-1.5\baselineskip}
    \centering
    \includegraphics[width=\textwidth]{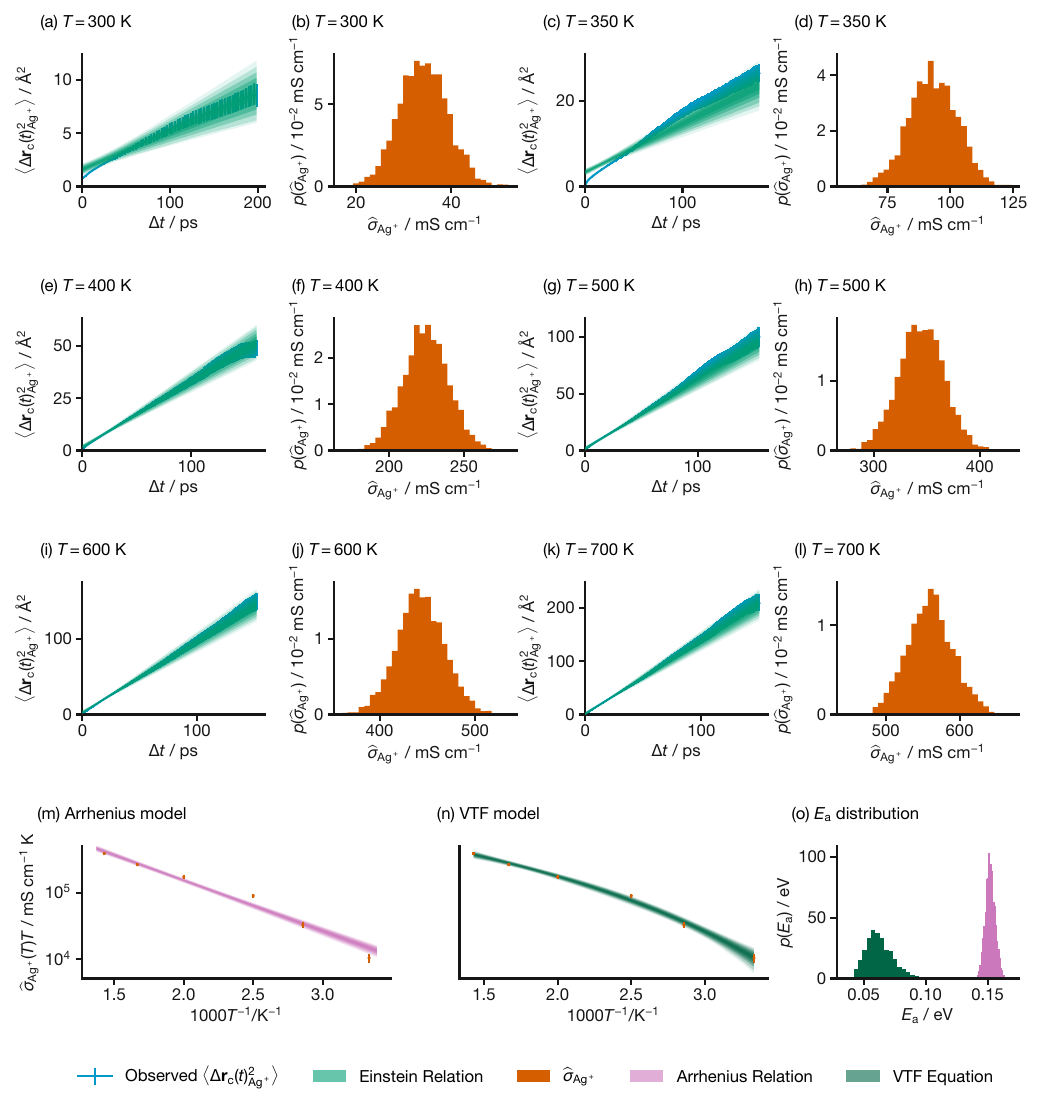}
    \vspace{-2\baselineskip}
    \script{simulation.py}
    \caption{The mean squared displacement data and associated $\sigma$ distributions at temperatures of \SIlist{300;350;400;500;600;700}{\kelvin} with \SI{130}{\pico\second} of diffusive simulation (a-l), the appropriate Arrhenius (m) and VTF model (n) plots and the resulting distributions of $\Ea$ from each modelling approach (o).}
    \label{fig:agcrse2_130}
    \vspace{-1.5\baselineskip}
\end{figure*}

\begin{figure*}
    \vspace{-1.5\baselineskip}
    \centering
    \includegraphics[width=\textwidth]{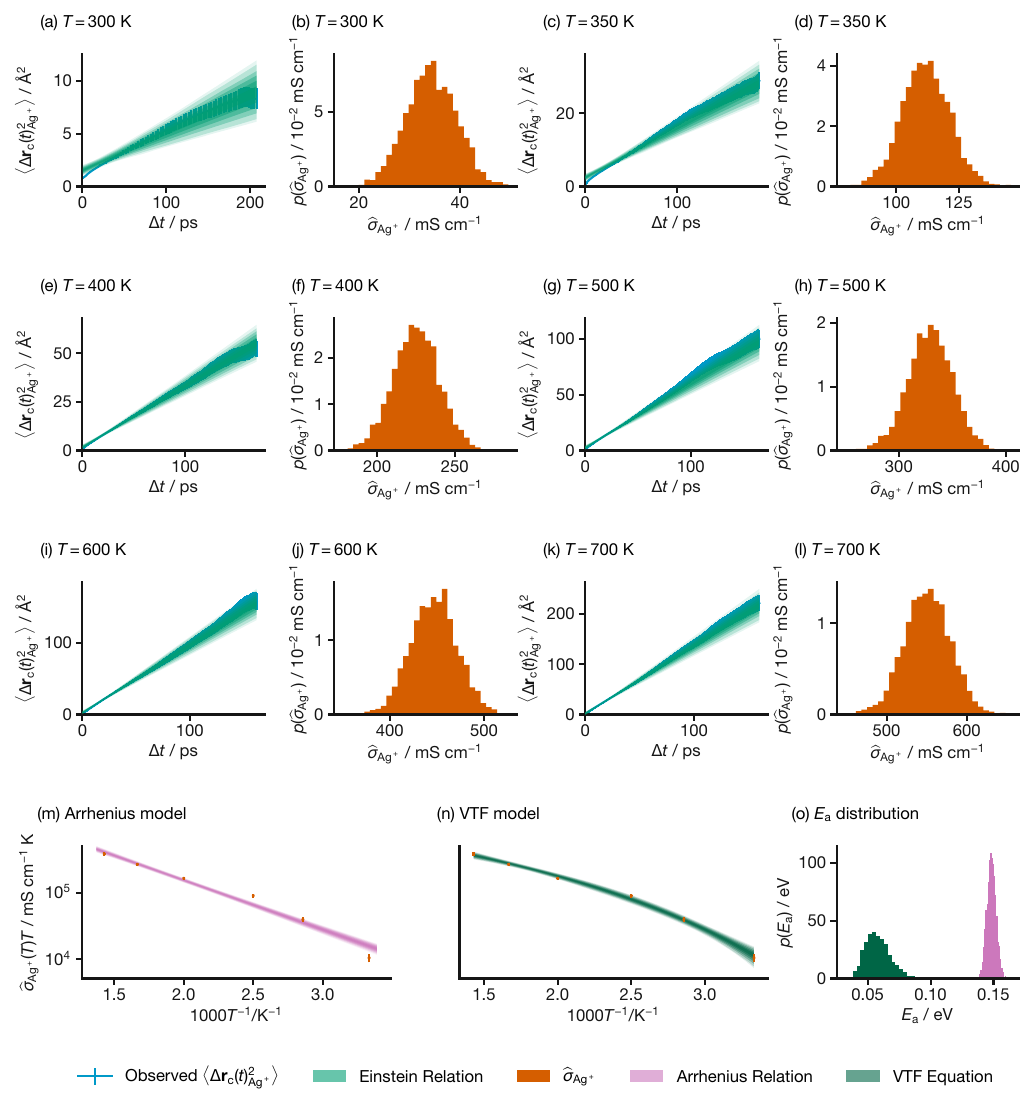}
    \vspace{-2\baselineskip}
    \script{simulation.py}
    \caption{The mean squared displacement data and associated $\sigma$ distributions at temperatures of \SIlist{300;350;400;500;600;700}{\kelvin} with \SI{140}{\pico\second} of diffusive simulation (a-l), the appropriate Arrhenius (m) and VTF model (n) plots and the resulting distributions of $\Ea$ from each modelling approach (o).}
    \label{fig:agcrse2_140}
    \vspace{-1.5\baselineskip}
\end{figure*}

Figs.~\ref{fig:agcrse2_40} to \ref{fig:agcrse2_140} show the mean-squared displacement as a function of lag time for temperatures of \SIlist{300;350;400;500;600;700}{\kelvin}, for increasing total simulation lengths.
Longer simulations provide more data in the diffusive regime, reducing the uncertainty in the conductivity estimates and, as discussed in Sec.~\ref{sec:model}, increasing the Bayesian evidence for the VTF model.


\end{document}